\documentclass[a4paper]{article}

\usepackage{INTERSPEECH2021}
\usepackage{color}
\usepackage{comment}
\usepackage{empheq}
\usepackage{multirow}
\usepackage{cite}
\usepackage{amsthm}

\newcommand{\ie}{\textit{i.e.,~}}
\newcommand{\eg}{\textit{e.g.,~}}

\newcommand{\etal}{\textit{et~al.~}}

\title{Manifold-Aware Deep Clustering: \\
Maximizing Angles between Embedding Vectors Based on Regular Simplex
}

\name{Keitaro Tanaka$^{\dagger *}$, Ryosuke Sawata$^\ddagger$, Shusuke Takahashi$^\ddagger$}
\address{
  $^\dagger$Waseda University, Japan, \hspace{5mm}
  $^\ddagger$Sony Corporation, Japan}
\email{phys.keitaro1227@ruri.waseda.jp, \{Ryosuke.Sawata, Shusuke.Takahashi\}@sony.com}

\begin{document}
\maketitle
\begin{abstract}
This paper presents
 a new deep clustering (DC) method 
 called manifold-aware DC (M-DC) 
 that can enhance hyperspace utilization more effectively than the original DC.
The original DC has a limitation in that a pair of two speakers has to be embedded 
 having an orthogonal relationship 
 due to its use of the one-hot vector-based loss function, 
 while our method derives a unique loss function
 aimed at maximizing the target angle in the hyperspace  
 based on the nature of a regular simplex.
Our proposed loss imposes a higher penalty than the original DC
 when the speaker is assigned incorrectly.
The change from DC to M-DC can be easily achieved 
 by rewriting just one term in the loss function of DC,
 without any other modifications to 
 the network architecture or model parameters.
As such, our method has high practicability 
 because it does not affect the original inference part.
The experimental results show
 that the proposed method improves the performances of the original DC and its expansion method.
\end{abstract}


\renewcommand{\thefootnote}{\fnsymbol{footnote}}
\footnote[0]{\scriptsize
$*$Work done during an internship at Sony.
Keitaro Tanaka wishes to express the deepest gratitude 
 to Prof. Shigeo Morishima 
 at Waseda Research Institute for Science and Engineering,
 Tokyo, Japan.
Keitaro Tanaka was fully supported 
 by JST Mirai Program No. JPMJMI19B2, 
 and JSPS KAKENHI Nos. 19H01129 and 19H04137 
 in the publication of this paper 
 and participation in the conference.
}

\vspace{-2mm}
\section{Introduction}

Monaural speech separation is a fundamental task
 for automatic speech recognition (ASR) \cite{ASR}. 
Before the widespread use of deep learning methods,
 most speech separation approaches focused on scenarios
 with multiple microphones \cite{nonDeep, DC_advance},
 and a monaural scenario
 remained a challenging task.
Since deep neural networks (DNNs) have appeared,
 a straightforward approach for this task has been
 to train a DNN
 in a supervised manner
 using pairs of mixture sound 
 and separated sounds \cite{early2}.
In these methods,
 one separation is generally performed per short duration 
 since such processing is convenient in terms of the calculation cost, simplicity to exploit, later application (\eg ASR) and so on.
However, this causes a permutation problem 
 that makes considering the consistency of output tracks difficult 
 because each DNN-based separation 
 is conducted per short duration independently.
For example, even if a temporary separation is successful,
 the speaker of the current 1st track is not always the same 
 as the speaker of the previous 1st track or the next 1st track,
 which may be swapped with the speaker of 2nd track.
There are two main approaches to avoiding this problem,
 and several DNN-based source separation methods using these approaches have been studied
 \cite{PIT1, PIT2, PIT3, DC1, DC2}.

The first approach involves
 permutation invariant training (PIT) \cite{PIT1, PIT2, PIT3}.
This method considers all possible correspondences
 between the outputs of a DNN
 and the speakers included in a mixture sound.
The actual optimization of the network is then conducted 
 only for the best permutation among them.
By considering all possible correspondences,
 PIT allows a DNN to be trained directly.
Furthermore,
 PIT works with any DNN output,
 including short-time Fourier transform (STFT) magnitudes
 \cite{PIT1, PIT2, PIT3} 
 and waveforms \cite{Conv-TasNet, DPRNN, nachmani20 ,wavesplit}.
Methods with waveform outputs
 have been particularly widely used
 because of their excellent quality
 (\eg Conv-TasNet \cite{Conv-TasNet}, 
      Dual-path RNN \cite{DPRNN},
      and Nachmani \etal \cite{nachmani20})
 and are currently the leading solutions
 for speaker-independent monaural speech separation.
Most recently, Zeghidour \etal have proposed Wavesplit \cite{wavesplit},
 which achieves a state-of-the-art performance.

The second approach is to use deep clustering (DC) \cite{DC1, DC2},
 which estimates a high-dimensional embedding
 for each time-frequency (TF) bin of an input spectrogram
 such that two embeddings are close to
 or far from each other 
 depending on whether they belong to the same speaker or not.
TF bins are grouped in accordance with speakers 
 by means of a clustering technique 
 (\eg {\it{k}}-means \cite{k-means})
 in the embedding space
 at a later stage.
Since the clustering technique is a non-supervised method 
 and can group arbitrary embeddings by speakers 
 regardless of the time duration, 
 DC can select a consistent track with the same speaker.
In this way,
 DC solves the permutation problem 
 by avoiding the need for direct mask estimation.
Inspired by this approach,
 many improved versions of DC
 have been proposed \cite{DC_advance}.
For example,
 the deep attractor network \cite{danet1, danet2, danet3}
 estimates not only embeddings
 but also the centroids of embeddings,
 and the Chimera network \cite{chimera1, chimera2}
 estimates both binary masks and ratio masks.
In addition, DC has been used
 in music information retrieval tasks,
 such as the Cerberus network \cite{cerberus} 
 and instrument-independent music transcription \cite{tanaka20},
 indicating its potential for various applications.

\begin{figure}[t]
  \centering
  \includegraphics[width=\linewidth]{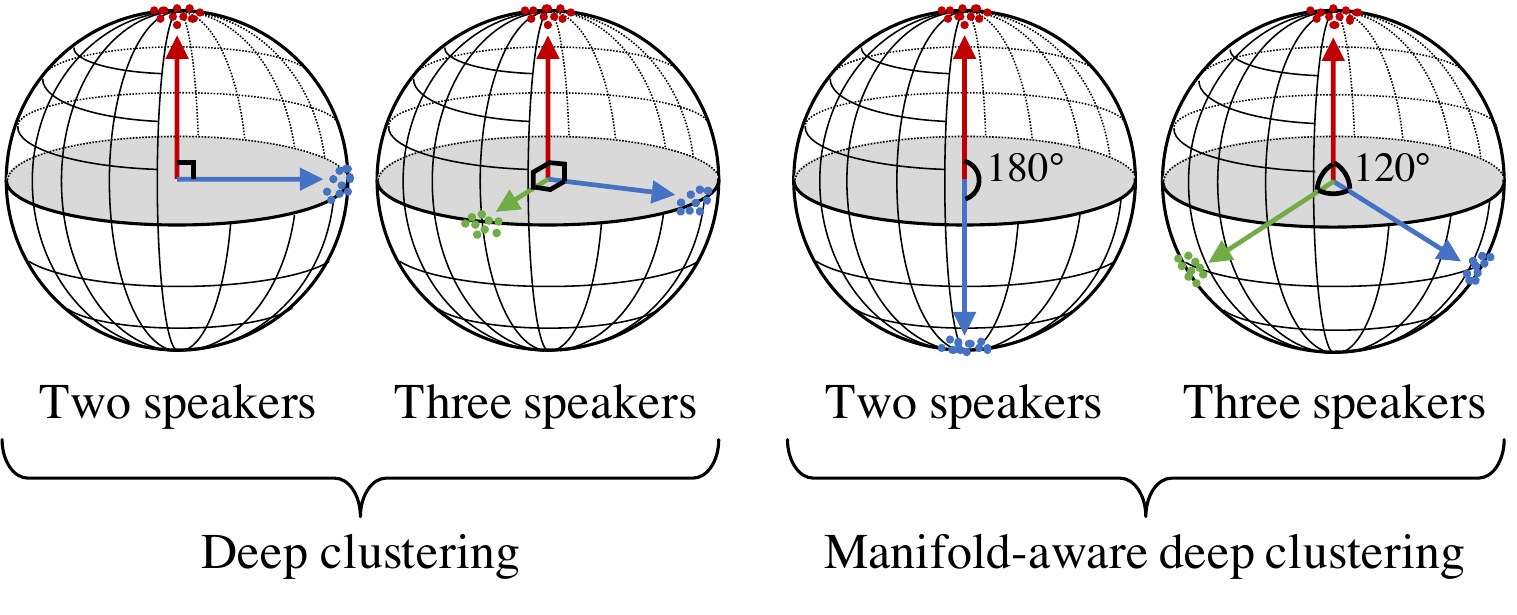}
  \vspace{-7.5mm}
  \caption{The proposed method maximizes the target angle 
            between two embeddings from different speakers.
           The target angle differs depending on the number of speakers.}
  \label{fig:concept}
  \vspace{-5.3mm}
\end{figure}

Despite the advances mentioned above,
 the nature of the embedding space in DC
 has not been studied sufficiently. 
In general, 
 the loss function of DC defines the \textit{correct} target distances
 between all pairs of embeddings of TF bins by using one-hot vectors.
In terms of the manifold perspective, 
 the embeddings belonging to different speakers
 are forced to locate
 in an orthogonal way
 on a hyperspherical space,
 since the representation of one-hot vectors corresponds to 90 degrees 
 by handling their vectors using a cosine function.
As an example,
 the vectors of DC are illustrated on the left side of Fig.~\ref{fig:concept}.
This orthogonal constraint causes inefficient use of the embedding space
 with respect to each input mixture,
 resulting in performance deterioration of DC.

In order to solve this problem,
 we propose a manifold-aware DC (M-DC)
 that renews the definition of target embedding distances.
To utilize the whole embedding space
 for each mixture,
 we maximize the target angle
 based on the number of speakers,
 as shown in Fig.~\ref{fig:concept}.
Specifically,
 we rewrite the original DC loss function
 to modify target cosine distances
 on a manifold hypersphere
 by considering the nature of a regular simplex
 \cite{rs1, rs2, rs3}.
In our method, 
 this newly derived cosine distance enables 
 M-DC to impose a smaller or greater penalty 
 to embeddings depending on whether they are easy or difficult to separate from already embedded speakers.
This effect
 leads to clustering-friendly embedding at the later stage.
Therefore,
 M-DC can discriminate the speakers more effectively
 than the original DC.

The main contribution of this paper
 is the operation of DC
 with a unique loss function
 for efficient learning
 based on full utilization of the hyperspherical embedding space with respect to each input mixture.
The change from DC to M-DC
 can be achieved 
 by rewriting just one term 
 in the loss function of DC,
 without any other modifications 
 to the network architecture 
 or model parameters.

\vspace{-1.5mm}
\section{Proposed Method}
Our goal with M-DC
 is to train the DC network
 while enhancing the utilization
 of its hyperspherical embedding space.
Specifically, 
 M-DC maximizes each of the angles between two speakers, 
 which are shaped in the embedding space.

\vspace{-1.5mm}
\subsection{Brief Review of Deep Clustering}
\vspace{-1mm}
Since our method is partly based on the original DC,
 we first review DC here briefly.

DC separates the input mixture speech
 in the STFT domain
 by assigning a $D$-dimensional normalized embedding
 to each TF bin
 using a DNN.
The network is trained 
 such that embeddings belonging to the same speaker
 are close to each other,
 and those belonging to different speakers 
 are far from each other.
Let $\bm{V} \!\! \in \! \mathbb{R}^{TF \times D}$ 
 and $\bm{Y} \!\! \in \! \{0,1\}^{TF \times N}$
 be the normalized embeddings estimated by a DNN
 and the dominant speaker indicators
 ($(\bm{Y})_{in}\!=\!1$ if speaker $n$ is dominant in the $i$th TF bin $(t,f)$,
  and $(\bm{Y})_{in}\!=\!0$ otherwise),
 where $T$, $F$, and $N$ respectively
 represent the numbers of time frames,
 frequency bins, 
 and speakers included in the mixture spectrogram.
DC brings the cosine distances between all embedding pairs
 close to the \textit{correct} target ones
 by minimizing the loss function $\mathcal{L}_{\mathrm{DC}}$,
 calculated as
    \begin{align}
    \mathcal{L}_{\mathrm{DC}} 
    &= \|\bm{V}\bm{V}^{\!\top}
          - \bm{Y}\bm{Y}^{\!\top}\|^2_F 
    \label{DC_loss}\\
    &= \|{\bm{V}^{\!\top}\bm{V}}\|^2_F
          + \|{\bm{Y}^{\!\top}\bm{Y}}\|^2_F
          - 2\|{\bm{V}^{\!\top}\bm{Y}}\|^2_F ,
    \label{DC_loss_exp}
    \end{align}
 where $\|\cdot\|_F$ denotes the Frobenius norm of a matrix.
$\bm{V}\bm{V}^{\!\top} \!\! \in \!\! \mathbb{R}^{TF \times TF}$
 shows the cosine distances between all embedding pairs,
 while $\bm{Y}\bm{Y}^{\!\top} \!\!\! \in \!\! \mathbb{R}^{TF \times TF}$
 shows the \textit{correct} target ones.
Note that $(\bm{Y}\bm{Y}^{\!\top})_{i_1i_2}\!\!=\!\!1$
 if the $i_1$th and $i_2$th TF bins
  belong to the same speaker,
  and $=\!0$ otherwise.
Since a DC network well-trained 
 by the above $\mathcal{L}_{\mathrm{DC}}$ 
 enables the embeddings to get close 
 to others obtained from the same speakers,
 assigning the TF bins to the correct speakers, 
 \ie clustering, becomes feasible.
 
\vspace{-1.5mm}
\subsection{Manifold-Aware Deep Clustering}
\vspace{-1mm}
In general, 
 to utilize the hyperspace effectively 
 with respect to each input mixture,
 it is desirable to divide the whole hypersphere into $N$,
 \ie the total number of speakers.
Specifically, 
 the hypersphere is divided by $N$ speakers equally
 so that each angle is equal to $\arccos \{-1 / (N\!-\!1)\}$.
For example, 
 each optimal angle is 180 degrees ($= \! \arccos \{-1 / (2\!-\!1)\}$) 
 if there are two speakers,
 and then its relationship between the embedded points 
 is equal to a straight line,
 which is a 1-dimensional regular simplex \cite{rs1, rs3}.
When there are three speakers,
 each optimal angle is 120 degrees ($= \! \arccos \{-1 / (3\!-\!1)\}$),
 and then its relationship among embedded points is equal to an equilateral triangle,
 which is a 2-dimensional regular simplex.
Similarly, such a relationship among four embedded speakers
 makes a regular tetrahedron,
 which is a 3-dimensional regular simplex.
On the basis of this discussion, 
 we inductively estimate that the embedded $N$ speakers make 
 an $(N\!-\!1)$-dimensional regular simplex.

Therefore, 
 to improve the hyperspace utilization of the original DC,
 the key thing to consider is the characteristics
 of the regular simplex.

\vspace{-2mm}
\subsubsection{Introduction of Regular Simplex}
\vspace{-1.5mm}

In order to enhance the utilization of a $D$-dimensional hyperspace,
 our goal is to determine
 the best $N$ points on a hypersphere
 (\ie a ($D\!-\!1$)-dimensional hypersphere)
 so that they are distributed as far away from each other as possible.
Note that
 the number of dimensions $D$ is appropriately large
 ($D \geq N-1$) for the given $N(\geq2)$.
Under the aforementioned condition, 
 we consider that the vertices of an ($N\!-\!1$)-dimensional regular simplex 
 are equivalent to the best $N$ positions,
 as described above.

First, 
 we show that the positional relationship we look for 
 needs to be that of the vertices of a regular simplex.
To ensure the symmetry among all points,
 they have to be located point-symmetrically
 with respect to the center of the hyperball.
Therefore, 
 the $D$-dimensional Euclidean distances between all pairs of the points
 are equal to each other.
In addition, 
 for $R \leq N-1$,
 none of the $R\!+\!1$ points can be contained 
 in an ($R\!-\!1$)-dimensional hyperplane at the same time.
This situation precisely
 meets the definition of a regular simplex.

Next, 
 we conversely show that 
 the positional relationship of the vertices of a regular simplex 
 can meet the relationship we look for.
An ($N\!-\!1$)-dimensional regular simplex
 can be embedded in a $D$-dimensional hyperball
 with a ($D\!-\!\!1$)-dimensional hypersphere 
 under the condition of $D \geq N-1$.
When $D>N-1$,
 the dimensions of $D\!-\!N\!+\!1$
 create
 extra dimensions
 during the embedding.
Due to the nature of the regular simplex,
 the vertices are necessarily point-symmetric with respect to the center
 and are located farthest away from each other.
Therefore, the two relationships are indeed equivalent.

{\tabcolsep=2.5mm
\begin{table*}[t]
\centering
\caption{Separation results of DC and M-DC.
         Scores written in bold are 
          the higher one(s).}
\vspace{-3mm}
\label{vsDC}
\begin{tabular}{l|cc|cc|cc|cc|cc}
\toprule
\multirow{2}{*}{Configuration}                                  & \multicolumn{2}{c}{SI-SDR} & \multicolumn{2}{c}{SDR} & \multicolumn{2}{c}{SIR} & \multicolumn{2}{c}{SAR} & \multicolumn{2}{c}{STOI} \\
& DC       & M-DC     & DC       & M-DC     & DC        & M-DC       & DC       & M-DC     & DC       & M-DC      \\ \midrule
2 speakers, 8 kHz     & 8.863      & \textbf{9.321}      & 9.402      & \textbf{9.843}    & 17.810       & \textbf{18.372}        & 10.473     & \textbf{10.837}      & 0.858      & \textbf{0.870}       \\
2 speakers, 16 kHz    & 6.923      & \textbf{7.347}      & 7.417      & \textbf{7.827}      & 15.923       & \textbf{16.405}        & 8.548      & \textbf{8.907}      & 0.810      & \textbf{0.822}       \\
3 speakers, 8 kHz     & 2.390      & \textbf{2.455}      & 3.271      & \textbf{3.331}      & 10.041       & \textbf{10.121}        & 5.376      & \textbf{5.423}      & 0.675      & \textbf{0.677}       \\
3 speakers, 16 kHz    & \textbf{0.889}      & 0.882      & \textbf{1.705}      & 1.701       & 8.629      & \textbf{8.645}        & \textbf{3.900}      & 3.874      & \textbf{0.611}      & \textbf{0.611}       \\ \bottomrule
\end{tabular}
\vspace{-1mm}
\end{table*}
}

\begin{figure*}[t]
  \centering
  \centerline{\includegraphics[width=0.91\linewidth]{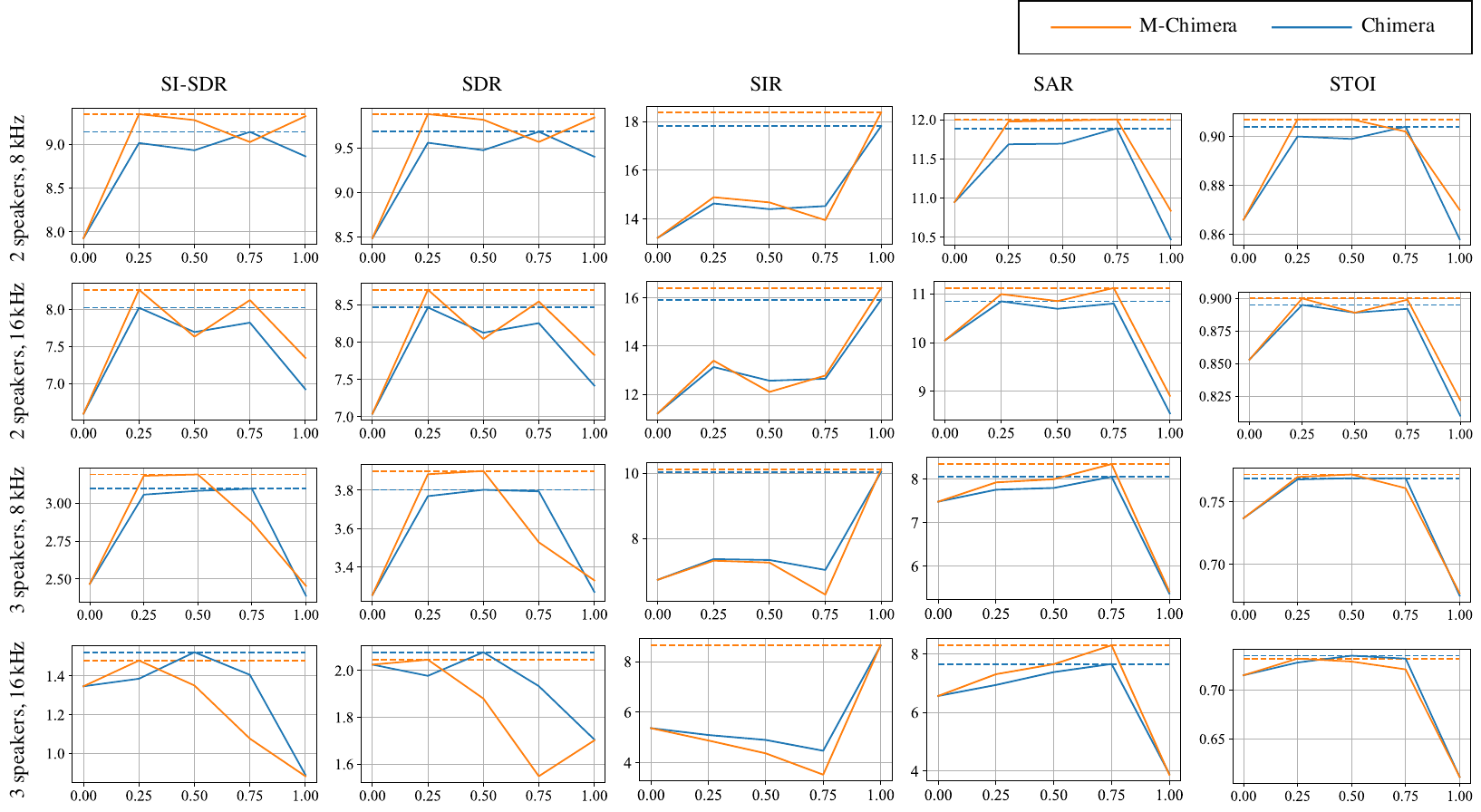}}
  \vspace{-2mm}
  \caption{Separation results of the Chimera network with DC (Chimera)
            and with M-DC (M-Chimera).
           Dashed lines show the best scores of each method
            under the configuration.
           The horizontal axis represents $\alpha$.
           M-Chimera is equivalent to Chimera when $\alpha\!=\!0$.}
  \label{fig:vsChimera}
  \vspace{-5mm}
\end{figure*}

\vspace{-2mm}
\subsubsection{New Definition of Target Cosine Distance}
\vspace{-1.5mm}

Motivated by the above discussion,
 we embed the mixed speakers as $N$ points
 in such a manner
 that the positional relationship meets the
 vertices of an ($N\!-\!1$)-dimensional regular simplex.
We can embed an ($N\!\!-\!\!1$)-dimensional regular simplex
 in a normalized hyperball
 by considering $N$ points $\{ {\bf x}_n \}_{n=1}^N$,
 where their coordinates from the centroid are represented as
    \begin{equation}
      (\underbrace{
       \hspace{0.3em} \cdots, \frac{-1}{N}\sqrt{\frac{N}{N\!\!-\!\!1}}
       }_{n-1},
       \frac{N\!\!-\!\!1}{N}\sqrt{\frac{N}{N\!\!-\!\!1}},
       \underbrace{
       \frac{-1}{N}\sqrt{\frac{N}{N\!\!-\!\!1}}, \cdots \hspace{0.3em}
       }_{N-n}) .
      \label{vertices}
    \end{equation}
Then, the inner products of pairs of these vectors
 are calculated as
    \begin{empheq}
        [left={{\langle {\bf x}_{n_1} \,, {\bf x}_{n_2} \rangle} =\empheqlbrace}]{alignat=2}
        & \hspace{0.3em} 1 
        && (n_1 = n_2)
        \label{inner_products1}
        \\
        &-\frac{1}{N-1} \hspace{1em} 
        && (n_1 \neq n_2) ,
        \label{inner_products2}
    \end{empheq}
 where $\langle\cdot,\cdot\rangle$ denotes the inner product of two vectors.
Equation~\eqref{inner_products1} ensures
 that the regular simplex just fits in a normalized hyperball
 (\ie the vertices touch a normalized hypersphere),
 while Eq.~\eqref{inner_products2} gives 
 the new target cosine distance.
Consequently, maximization of the angle corresponding to this cosine distance, 
 \ie $\arccos \{-1 / (N\!-\!1)\}$, becomes feasible.

Equations~\eqref{vertices},~\eqref{inner_products1},~and~\eqref{inner_products2}
 also show
 that M-DC gets closer to the original DC
 when $N$ increases.
M-DC theoretically matches the original DC
 in a situation with
 an infinite number of speakers $N$, as follows:
    \begin{align}
    &{\rm{Eq.}~\eqref{vertices}} \xrightarrow{N\rightarrow\infty}
      (\underbrace{
       \hspace{0.3em} \cdots, 0
       }_{n-1}, 
       1,
       \underbrace{
       0, \cdots \hspace{0.3em}
       }_{N-n})
      \label{vertices_inf} , \\
    &{\rm{Eq.}~\eqref{inner_products1}} \xrightarrow{N\rightarrow\infty}
      1 \: (=\cos (0))
      \label{inner_products1_inf} , \\
    &{\rm{Eq.}~\eqref{inner_products2}} \xrightarrow{N\rightarrow\infty}
      0 \: (=\cos (90))
      \label{inner_products2_inf} .
    \end{align}
As shown in Eqs.~\eqref{inner_products1_inf} and \eqref{inner_products2_inf}, 
 each of the angles
 among all embedded points is 90 degrees, \ie orthogonal, 
 when assuming the total number of speakers is infinity, 
 and this is equal to the original DC's loss 
 due to using one-hot vector,
 like Eq.~\eqref{vertices_inf}.
Hence, M-DC is equal to the original DC when the total number of speakers $N$ is infinity.
In other words, M-DC is theoretically equivalent or superior to the original DC since $N$ is actually a finite value.

\vspace{-1.0mm}
\subsubsection{Manifold-Aware Deep Clustering Training}
\vspace{-1.0mm}
M-DC not only maximizes the angle 
 between two embeddings belonging to different speakers 
 but also allows the use
 of previously untouched sides of the hypersphere
 due to its overall symmetry.
This theoretically enables us
 to train a DNN having maximum spatial efficiency for each mixture.
Furthermore, the only difference between DC and M-DC
 in the actual training is exactly
 one term of $\mathcal{L}_{\mathrm{DC}}$
 (\ie $\bm{Y}$, not $\bm{Y}\bm{Y}^{\!\top}$).
Since it is difficult
 to calculate $\mathcal{L}_{\mathrm{DC}}$
 directly defined in Eq.~\eqref{DC_loss}
 because of its high computational cost,
$\mathcal{L}_{\mathrm{DC}}$ is actually calculated
 with its expanded form (Eq.~\eqref{DC_loss_exp}).
On the basis of Eqs.~\eqref{vertices}--\eqref{inner_products2},
 we can define the single term $\bm{Y}$ for M-DC as follows:
    \begin{equation}
      \bm{Y} = (
      \underbrace{\hspace{0.3em} \cdots, {\bf y}_i,\cdots \hspace{0.3em} 
      }_{TF})^{\!\top} ,
      \label{singleY}
    \end{equation}
 where ${\bf y}_i = {\bf x}^{\!\top}_n$
 if speaker $n$ is dominant in the $i$th TF bin.

In summary,
 substituting our new $\bm{Y}$ defined 
 in Eq.~\eqref{singleY} for Eq.~\eqref{DC_loss_exp} 
 means we can easily achieve M-DC.
Hence, changing M-DC from the original DC becomes feasible 
 without any other modifications 
 to the network architecture or model parameters.
\section{Evaluation}

In this section,
 we evaluate the proposed method by replacing the original DC 
 in two well-known methods with our M-DC:
 i) the original DC itself and ii) the DC-based Chimera network.
 
\vspace{-1.5mm}
\subsection{Data}
\vspace{-1.5mm}
In our evaluation,
 we used the WSJ0-2mix and WSJ0-3mix datasets \cite{DC1},
 which contain 30 hours of 
 two- and three-speaker speech mixtures for training 
 and ten hours for validation.
The evaluation sets contain 
 five hours of two- and three- speaker speech mixtures, 
 where the speakers are not shared 
 between the training or the validation sets 
 and the evaluation sets. 
All sounds were resampled to 8 kHz and 16 kHz.
We used STFT with the square root of a Hann window
 of 256 samples and 
 a shifting interval of 64 samples,
 \ie 75\% overlapped,
 to obtain the spectrogram.

\vspace{-1mm}
\subsection{Model Configurations}
\vspace{-1mm}
The DC network
 had four bi-directional long short-term memory (BLSTM) layers
 followed by one feedforward layer.
Each BLSTM layer had 600 hidden cells,
 and the feedforward layer corresponded
 to the embedding dimensions,
 which we set to $D\!=\!40$.
The Chimera network
 had a similar architecture
 except that the feedforward layer was split into two.
One was the same as the DC network (DC head),
 while the other corresponded
 to the number of speakers (MI head)
 for direct mask inference.
We used rmsprop optimization \cite{rmsprop}
 with an initial learning rate of $10^{-4}$.
We dropped the learning rate by 0.5
 if the validation loss
 did not decrease after five epochs.
We stopped the training 
 if the validation loss
 did not decrease after thirty epochs
 and then selected the network
 on the basis of the validation loss.
The batch size was 32.
Our implementation used Asteroid \cite{asteroid},
 the PyTorch-based audio source separation toolkit.

\vspace{-1mm}
\subsection{Evaluation Criteria}
\vspace{-1mm}
We evaluated M-DC
 in terms of five criteria:
 scale-invariant signal-to-distortion ratio (SI-SDR) in dB \cite{SI-SDR},
 signal-to-distortion ratio (SDR) in dB \cite{SDR_SAR_SIR1},
 signal-to-interference ratio (SIR) in dB \cite{SDR_SAR_SIR1},
 signal-to-artifacts ratio (SAR) in dB \cite{SDR_SAR_SIR1},
 and short-time objective intelligibility (STOI) ranging from 0 to 1 \cite{STOI}.
We calculated the averages
 of the above criteria for each speaker.
The performance on DC was evaluated
 with a different number of speakers (two or three)
 and different sampling rates (8 kHz or 16 kHz).
Clustering at a later stage
 was conducted by {\it{k}}-means.
The performance on the Chimera network 
 was additionally evaluated
 for different hyperparameter values 
 $\alpha \! \in \! \{0, 0.25, 0.5, 0.75, 1.0 \}$,
 which controls the weight between the two heads
 (only MI head or DC head was used 
  when $\alpha\!=\!0$ or $\alpha\!=\!1$).
The sound separation was conducted
 by the MI head
 (\ie by estimated ratio masks)
 except in the case of $\alpha\!=\!1$ 
 (\ie by estimated binary masks).
 
\vspace{-1mm}
\subsection{Experimental Results}
\vspace{-1mm}


The experimental results are shown in Table~\ref{vsDC} and Fig.~\ref{fig:vsChimera}.
First, Table~\ref{vsDC} shows that our M-DC outperformed the original DC 
 in almost all cases.
In particular, our performances regarding two speakers 
 were significantly superior to those of DC.
While we found that 
 the degrees of improvement in the case of three speakers were less than those of two speakers, 
 they were comparable to the original DC's ones.
This is because, as we explained in Sec.~\textbf{2.2.2} 
 (see Eqs.~\eqref{vertices_inf}--\eqref{inner_products2_inf}),
 the performance of M-DC got close to that of the original DC
 by increasing the number of speakers $N$, \ie from 2 to 3, in these experiments.
Hence, our assumption regarding the relationship 
 between the original DC and our M-DC described in Sec.~\textbf{2.2.2} is correct.

Next, 
 we focus on the results of the DC-based expansion method, 
 \ie the Chimera network.
From Fig.~\ref{fig:vsChimera},
 we can see that
 almost all our best scores were superior to those of the original Chimera network 
 (as indicated by the dashed line).
In addition, 
 we confirmed a tendency that was similar to the previous results, 
 \ie M-DC getting close to DC by increasing the number of speakers.

\begin{figure}[t]
  \centering
  \centerline{\includegraphics[width=0.79\linewidth]{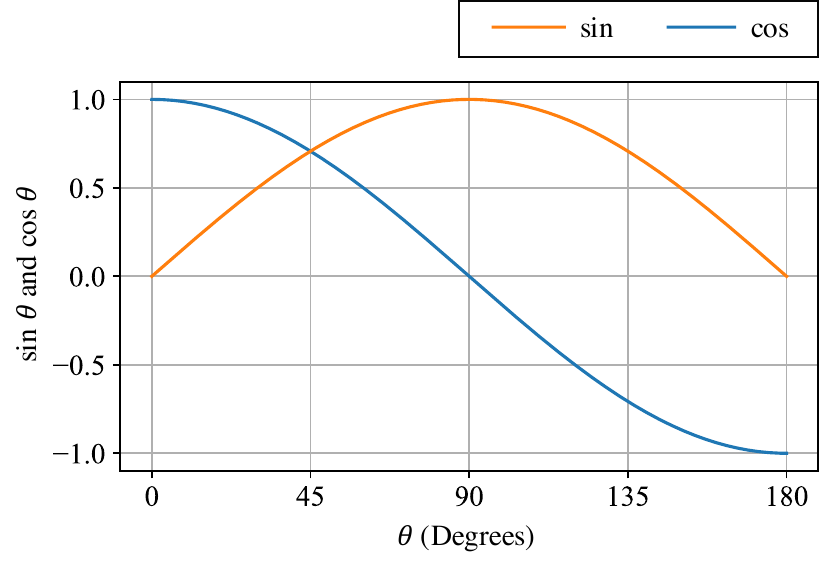}}
  \vspace{-4mm}
  \caption{The sine represents 
            the absolute value of the gradient of the cosine,
            which reaches its maximum at 90 degrees.}
  \label{fig:sin_vs_cos}
  \vspace{-5.5mm}
\end{figure}

Here, 
 we focus on the difference of the optimal $\alpha$
 between DC and M-DC loss.
Since both methods learn based on the cosine distance,
 we presume that the training depends on the value of the cosine's gradient.
Specifically, a large penalty is imposed
 to the angle when the absolute value of the gradient of the cosine 
 (\ie the value of the sine)
 is large, as shown in Fig.~\ref{fig:sin_vs_cos}.
This suggests that the orthogonal position, \ie $\sin (90)$, 
 has the highest influence on the training 
 due to having the highest value in Fig.~\ref{fig:sin_vs_cos}.
As shown in Fig.~\ref{fig:concept}, in DC, the orthogonal position is equal to the different speaker.
On the other hand, in M-DC, the orthogonal position is equal to the ambiguous speaker 
 (not the different one) since our M-DC locates a different speaker having more than 90 degrees, \eg 180 and 120 degrees when the number of speakers $N$ is 2 and 3.
In other words, the orthogonal position in our M-DC means the ambiguous speaker,
 since the 90 degrees is under both 180 and 120 degrees, 
 and thus our M-DC was trained by focusing more
 on the difficult ($=$ ambiguous) cases than the original DC.
These results demonstrate
 that our M-DC can deliver a more robust 
 and effective performance than the original DC. 
This effect also 
 resulted in the difference of the best $\alpha$
 in the Chimera network.

\section{Conclusion}
In this paper, we presented M-DC,
 which improves the original DC by using
 a unique loss function
 for efficient learning.
We fully utilized
 the hyperspherical embedding space
 by newly defining a target embedding distance
 by means of a loss function
 depending on the number of speakers.
The change from DC to M-DC was easily achieved
 by modifying just one term of the loss function,
 and thus has no effect on the original DC's inference part.
Furthermore, since not only the original DC but also the DC-based expansion methods can be improved by introducing our M-DC,
 as demonstrated in our experiments, we argue that M-DC has high practicability for many existing DC-based methods.

In future work, 
 we will evaluate the performance improvement 
 based on the clustering accuracy itself.
We also plan to conduct additional evaluations on cases 
 of more than three speakers
 and in other DC-based techniques.
In this work, 
 we let M-DC create extra dimensions,
 and although we think these dimensions enrich a DNN's expression ability,
 there is a room for making the model more compact 
 by means of dimensional reduction.
Another interesting direction
 would be to combine our method
 with other clustering techniques \cite{FutureWork},
 which would result in an even better performance.
We believe the findings of our work
 will contribute to the further development of DC.

\clearpage

\bibliographystyle{IEEEtran}

\bibliography{mybib}

\end{document}